\documentstyle[psfig]{mn}

\begin{document}

\title{Two nearby M dwarf binaries from 2MASS}

\author[J.E. Gizis et al.]{John E. Gizis$^1$, 
David G. Monet$^2$,
I. Neill Reid$^3$,
J. Davy Kirkpatrick$^4$ and \newauthor 
Adam J. Burgasser$^5$ \\
$^1$Department of Physics and Astronomy, University of Massachusetts,
Amherst, MA 01003  \\
$^2$U.S. Naval Observatory, P.O. Box 1149, Flagstaff, AZ 86002 \\
$^3$Department of  Physics and Astronomy, University of Pennsylvania,
209 South 33rd Street, Philadelphia PA 19104-6396 \\
$^4$Infrared Processing and Analysis Center, 100-22, 
  California Institute of Technology,
  Pasadena, California 91125 \\
$^5$Department of Physics, 103-33, California Institute of Technology, 
Pasadena, California 91125
}

\maketitle

\begin{abstract}
We report the discovery of two binary M dwarf systems in the
immediate solar neighborhood using 2MASS.  The first is an M6.5 companion to
the nearby G star HD 86728 (Gl 376).  The known properties
of HD 86728 indicate that the M dwarf (Gl 376B) is old, metal-rich
and only 14.9 parsecs away.  The M dwarf is highly active with
both H$\alpha$ and X-ray emission.  Thus, Gl 376B offers the
opportunity to study an old, bright, active M dwarf with known metallicity,
age, and luminosity.  
We show that it is probable that Gl 376B is itself an
unresolved pair.  
The other system consists of an M6.5 and an M8 dwarf with 
14.5 arcseconds separation.  
We estimate a distance of $\sim 16$ parsecs
for this very low mass pair.  Stronger activity is observed in the
M6.5 dwarf, supporting evidence that chromospheric activity is
weakening near the hydrogen burning limit.  
\end{abstract}


\section{Introduction \label{intro}}

Nearby binary stars have considerable importance for 
astronomical studies.  The nearby star sample is
the basis for determining the properties of the
field disk population, such as the luminosity and mass functions,
kinematics, binary fraction, and chromospheric and coronal activity.
Furthemore, binary systems offer special opportunities
to astronomers.
Since the components of a system can be assumed to have
the same age and composition, studies of one component
can be used to constrain the properties of the other
component.  This is especially important for the coolest M dwarfs,
whose complicated atmospheres are difficult to model.  G dwarf
primaries allow many otherwise unmeasurable properties of
their M dwarf secondaries to be constrained, but surprisingly
few wide systems suitable for ground-based follow-up are known.

We have recently begun a search using the Two Micron All Sky Survey
(2MASS) aimed at identifying the nearest isolated M and L dwarfs in
the solar neighborhood.  Here we report the discovery of
two nearby wide M dwarf secondaries of special interest 
in the course of that search.  

\section{Data}

Candidate nearby M dwarfs were identified using the pairing of 2MASS 
detections with USNO PMM scans (Monet, private comm.) of 
the Second Palomar Sky Survey (POSSII) plates \cite{poss2}.  
On the basis of the resulting BRIJHK magnitudes, nearby
M dwarfs can be identified with high confidence.  More details
are given in Gizis et al. (1999, in prep.).  
The selection criteria are designed
to produce a complete sample of M8 and cooler dwarfs, but necessarily include
many M6.5 and M7 dwarfs.  Objects with $K_s<12$, $J-K_s>1.0$
were selected for initial followup.  

In the course of compiling the nearby star candidate list, 
we noted that two of the M dwarfs appeared to be companions to
brighter stars and marked them for priority in follow-up spectroscopy.
The 2MASS J2000 positions and magnitudes are 
given in Table~\ref{table-data}.  The first, 2MASSW J1000503+315545, 
is close (133 arcseconds east, 20 arcseconds north) to the nearby G star 
HD 86728, which appears in the
nearby star catalog as Gl 376 \cite{gj91}.  Measurement of the 
POSSI and POSSII plates reveals the 2MASS source's proper motion is 
identical to Gl 376.  
Indeed, it is a little surprising that this source has not been 
noticed before.  It is easily visible on the Palomar
Sky Survey plates, and its proper motion is sufficient to
be included in the LHS Catalogue (Luyten 1979), although it is
within the halo of the bright star.  
We estimate $R \approx 14.5 \pm 0.5$
from the uncalibrated PMM scans.  The common
proper motion indicates it is associated, and we 
therefore denote this object Gl 376B.  We henceforth refer to the
primary (HD 86728) as Gl 376A.   

\begin{figure}
\label{fig-spec}
\psfig{figure=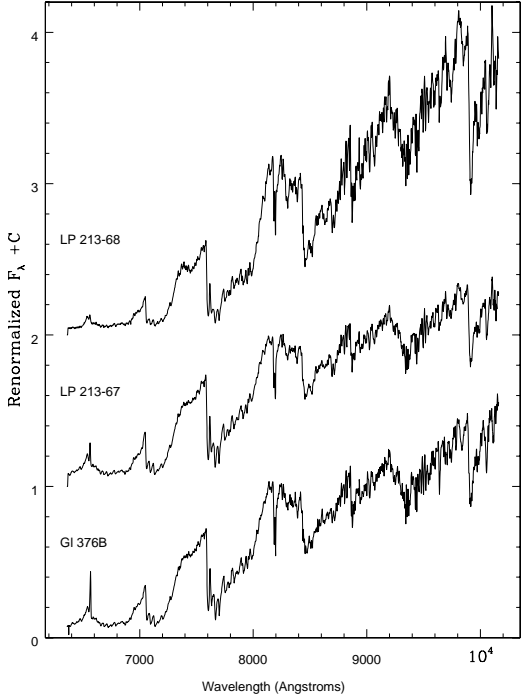,height=10cm}
\caption{The Keck spectra of the three nearby M dwarfs.}
\end{figure}

A spectrum of Gl 376B was obtained on 04 March 1999 
with LRIS spectrograph \cite{lris} on the Keck II telescope
using the setup described in Kirkpatrick et al. (1999).
A second spectrum was obtained at the Hale 200 in. telescope
at Palomar Mountain on 24 May 1999.  
The Keck spectrum is shown in Figure~\ref{fig-spec}.
The spectral type of Gl 376B is M6.5 V with an H$\alpha$ equivalent
width of 16\AA ~in the Keck spectrum and and 13 \AA~ in the lower resolution
Palomar spectrum.  Spectral types are on the Kirkpatrick, Henry \&
McCarthy (1991) system, and were determined by visually comparing our spectrum
to stars of known spectral type taken with the same setup.  

Our second system consists of the M dwarfs 
2MASSW J1047126+402643 and 2MASSW J1047138+402649.  
The two sources show common proper motion of $\mu_\alpha = -0.30$
and $\mu_\delta = -0.03$ arcseconds per year relative to other
stars in the field, as measured from the POSSI and 2MASS images,
and have a separation of 14.5 arcseconds.  
Both the position and the proper motion agree with the two
Luyten \shortcite{nltt} NLTT sources LP 213-67 and LP 213-68,
and we adopt those names henceforth.  
The two M dwarfs were observed at Keck II on 05 March 1999.  The primary,
LP 213-67, has a spectral type of M6.5 V and
an H$\alpha$ equivalent with of 6.9\AA.  The secondary,
LP 213-68, is an M8 V with an H$\alpha$ equivalent width of  
3.7 \AA.  

\begin{table*}
\begin{minipage}{200mm}
\caption{2MASS Data}
\label{table-data}
\begin{tabular}{crrrrrl}
{Name} &
{R.A.} &
{Dec.} &
{J} &
{H} &
{K$_s$} &
{Epoch} 
\\
Gl 376B & 10:00:50.30 & +31:55:45.9 & 10.30 & 9.68 & 9.27 & 21 March 1998 \\
LP 213-67 & 10:47:12.65 & +40:26:43.7 & 11.42 & 10.78 & 10.40 & 05 April 1998 \\ 
LP 213-68 & 10:47:13.82 & +40:26:49.3 & 12.45 & 11.71 & 11.28 & 05 April 1998 \\
\end{tabular}
\end{minipage}
\end{table*}

\section{Discussion}

Gl 376A has been extensively studied, and as a result, its properties
are well determined.  The Hipparcos parallax places it at
14.9 parsecs \cite{hipparcos}.  Taylor's (1995) analysis of all metallicity
determinations of the star indicates that $[Fe/H] = +0.170 \pm 0.052$. 
Using the Hipparcos distance and 
Edvardsson et al.'s (1993) photometry, Ng \& Bertelli (1998)
find $M = 1.02$ M$_\odot$ and $\log age = 9.854 \pm 0.083$.
This age is in good agreement with Barry's (1988) estimate
of $\log age = 9.93$ Gyr based on the CaII chromospheric emission lines.  
Duquennoy, Mayor \& Halbwachs (1991) find a constant radial velocity of
+55.83 km/s, and both the Hipparcos observations \cite{hipparcos} 
and ground based astrometry \cite{heintz} indicate that there is
no close companion.  Although there is an IDS and CCDM entry for 
a bright, nearby companion to Gl 376A in SIMBAD, this entry has been 
deleted from
the more recent Washington Visual Double Star Catalog
\cite{wds} and there is clearly no such double on 
the Palomar plates.  We note that Gl 376B should have been included
in the Duquennoy \& Mayor (1991) analysis of the frequency of
binary companions, though it appears in a separation range in
which they assumed 50\% incompleteness and therefore, in a sense,
was accounted for by their analysis.  

The Hipparcos parallax and 2MASS photometry implies $M_K=8.19$
for Gl 376B.  However, the typical absolute magnitude of an M6.5 
dwarf is $M_K = 9.60$ \cite{km94}.  
Furthermore, our estimate of $R \approx 14.5$ 
implies $M_R \approx 13.4$, which is 
brighter than the typical M6.5 value of $M_R=15.13$
\cite{km94}, but consistent with the $K_s$ discrepency.  
The apparent overluminosity of Gl 376B may be partially explained if
it is actually a near-equal luminosity binary (or triple!).  
The high metallicity of the system may also partially account for 
the apparent overluminosity of the M dwarf with respect to the 
typical field main sequence.

We may compare Gl 376B's H$\alpha$ emission strength to 
similar known M dwarfs observed by Hawley, Gizis \& Reid (1996).  
Despite its great age, Gl 376B is one of the most active
nearby field M dwarfs.  Since strong emission was observed twice,
it is unlikely that a flare is responsible.   
We note that the ROSAT All-Sky Survey \cite{1rxs} source
1RXSJ 100050.9+315555
is coincident with Gl 376B.
Adopting the calibration given by Fleming et al. (1993), we
find $f_X = 1.4 \times 10^{-12}$ erg s$^{-1}$ cm$^{-2}$ if the
M dwarf is the X-ray source.  With an
implied luminosity of $\log L_X = 27.5$ (or 27.2 if
it is an equal-luminosity binary), Gl 376B is very active
compared to the limited sample of late M dwarfs 
observed by Fleming et al.
While we cannot rule out the possibility that some of the
flux is due to the G primary, the ROSAT position with its
30 arcsecond uncertainty does not
match the G star and Barry \shortcite{barry} find little
chromospheric activity, which suggests that much or all of the observed
X-ray.  H\"{u}nsch et al. \shortcite{hssv99} 
find nearby old G stars with $\log L_X < 27$, 
and they do not attribute the ROSAT source to Gl 376A.  

The high activity level cannot be
due to youth, since the G primary is known to be old
from two independent techniques.  This strongly suggests
that Gl 376B is a short-period binary which maintains
a rapid rotation rate via tidal interaction \cite{ysh87}.  
An unresolved companion would thus explain both
the high activity level and the apparent overluminosity of 
Gl 376B.  
The postulated companion, Gl 376C, should be detectable via 
high resolution spectroscopy and radial velocity monitoring.

Both  members of the LP 213-67/68
system show H$\alpha$ emission.  
We note that the M6.5 primary has stronger emission than the M8 secondary.  
This is consistent with the general
observation that in the Pleiades and the field,
the chromospheric activity levels are becoming weaker 
as the hydrogen burning limit is approached \cite{tr98}, 
although the hotter
component could be more active due to chance or the influence of
an unseen companion.  

We have derived the relationship $M_K = 7.59 + 2.25 \times(J-K)$
using trigonometric parallax data for M7 and later dwarfs
(Gizis et al. 1999, in prep.).  
This implies a distance of 16 parsecs for the M8 secondary.
Another estimate may be obtained using Kirkpatrick \& McCarthy's (1993) 
value of $M_K=9.6$ for M6.5 dwarfs.
In this case, the estimated distance for the primary is 
14.5 parsecs.  The tangential velocity of $\sim 23$ km s$^{-1}$ 
does not a give a strong clue to the age of this system, but
an age of a few Gigayears is likely given the M6.5's activity level.  

\section{Conclusion}

We have discovered an M dwarf member of the Gl 376 system. 
This offers the opportunity to study a old, metal-rich 
M dwarf with known properties.  Most active, metal-rich objects
are young (such as the Hyades) and therefore it is difficult
to disentangle the effects of youth and metallicity.  
It is perhaps unfortunate that Gl 376B is likely an unresolved
double, because we cannot presently use its position 
to measure the shift of the main sequence with metallicity.  
On the other hand, detection of Gl 376C would allow 
mass ratio for the M dwarf to be determined, giving an
additional contraint for models, and would allow the
luminosity ratio of the system to be measured. 
Of course, if a direct orbital mass determination proves to be
possible, or if the system proves to be eclipsing, this
system would provide a unique constraint on stellar structure, 
atmospheric, and coronal models near the hydrogen
burning limit.  

We have also identified a nearby pair of cool M dwarfs.
LP 213-67 and LP 213-68 appear to be a normal
pair of M dwarfs just above the hydrogen burning limit.  
We note that the two components are easily resolved and would
have simultaneously measureable parallaxes.  With a separation of
$\sim 230$ AU, an orbital period is not likely in the near future,
but the system should allow comparative study of M6.5 and M8 dwarfs.

\section*{acknowledgments}

We thank Hartmut Jahrei{\ss} for discussion of the nature of
Gl 376A and James Liebert for comments.  
This work was funded in part by NASA grant AST-9317456
and JPL contract 960847.  
JDK and AJB acknowledge the support of JPL,
Caltech, which is operated under contract
with NASA.  INR and JDK acknowledge funding through
a NASA/JPL grant to 2MASS Core Project science; AJB acknowledges
support through this grant.  
This publication makes use of data products from 2MASS, 
which is a joint project of the
University of Massachusetts and IPAC, funded by NASA and NSF.
Some of the data presented herein were obtained at the W.M. Keck
Observatory, which is operated as a scientific partnership among 
the Caltech, the University of California 
and NASA.  The Observatory was made possible by the generous 
financial support of the W.M. Keck Foundation. 
This research has made use of the Simbad database, operated at
CDS, Strasbourg, France.


\end{document}